\documentclass[aps,pre,reprint,a4paper,showpacs]{revtex4-1}

\usepackage[T1]{fontenc}
\usepackage[latin9]{inputenc}
\usepackage{amsmath}
\usepackage{amssymb}
\usepackage{amsfonts}
\usepackage{graphicx}
\usepackage[english]{babel}
\usepackage{pifont}
\usepackage{natbib}
\usepackage{geometry}
\usepackage{txfonts}
\usepackage{hyperref}

\begin{document}

\title{Numerical solution of the stationary multicomponent nonlinear
Schr\"{o}dinger equation with a constraint on the angular momentum}

\author{Patrik Sandin}
\email{patrik.sandin@oru.se}
 
\author{Magnus \"{O}gren}
\author{M{\aa}rten Gulliksson}

\affiliation{School of Science and Technology, \"{O}rebro University, 70182 \"{O}rebro,
Sweden}

\date{\today}

\begin{abstract}
We formulate a damped oscillating particle method to solve the stationary nonlinear Schr\"{o}dinger equation (NLSE). The ground state solutions are found by a converging damped oscillating evolution equation that can be discretized with symplectic numerical techniques. The method is demonstrated for three different cases: for the single-component NLSE with an attractive self-interaction, for the single-component NLSE with a repulsive self interaction and a constraint on the angular momentum, and for the two-component NLSE with a constraint on the total angular momentum. We reproduce the so called yrast curve for the single-component case, described in [A.~D.~Jackson \emph{et al.}, Europhys. Lett. {\bf 95}, 30002 (2011)], and produce for the first time an analogous curve for the two-component NLSE. The numerical results are compared with analytic solutions and competing numerical methods. Our method is well suited to handle a large class of equations and can easily be adapted to further constraints and components.
\end{abstract}

\maketitle

\section{Introduction}

The nonlinear Schr\"{o}dinger equation (NLSE) is important in many different
fields of physics \cite{Fibich2015}: for example, in nonlinear optics \cite{AbdullaevOL2011};
in superconductivity that can be modeled with the related Ginzburg-Landau
equation \cite{OgrenPhysicaC2012}; in vortex line models for dual
strings in research in gravitation \cite{NielsenNuclPhysB1973}; and self-gravitating models for dark matter \cite{PhysRevD.84.043531}. Here we present examples of NLSEs in the settings of a mean-field
description of bosonic atoms, which has been an active area of research
since the experimental breakthrough in the mid-1990s when Bose-Einstein
condensates (BECs) were created in the laboratory with ultracold atomic
gases \cite{AndersenScience1995,BradleyPRL1995,DavisPRL1995,EnsherPRL1996}.
In this context the stationary NLSE typically has the form
of a Gross-Pitaevskii equation

\begin{equation}
-\frac{\hbar^{2}}{2M}\nabla^{2}\varphi+U_{0}\left|\varphi\right|^{2}\varphi=\mu\varphi,\label{eq:GPE}
\end{equation}
which is constrained by the normalization condition 
$\int \left|\varphi\right|^{2}\mathrm{d}V=N_\varphi$,
where $N_\varphi$ is the number of atoms, M is the mass of an atom, and the
atom-atom mean-field interaction parameter $U_{0}$ can be varied in sign and
amplitude, for example, by an external magnetic field.

After the experimental achievements of creating BECs in the laboratory, numerical modeling of various properties of condensates accelerated. For example different techniques to solve Eq. (\ref{eq:GPE}) \cite{DalfovoJRNIST1996,BaoJoCP2006,MalloryPRE2015,PhysRevE.91.053304}, as well as for the corresponding time dependent equation \cite{TahaJoCP1984} have been developed. 
For modeling BEC dynamics it is crucial to maintain the normalization condition; see, e.g., \cite{CerimeleCPC2000,CerimelePRE2000} for methods that fulfill this to machine precision in each time step. 

Furthermore, methods using a quantum lattice Boltzmann equation have been proposed to model expanding condensates \cite{PalpacelliPRE2008} over long times.
More recently a connection between the Kohn-Sham equations, that can be used in density functional theory of bosonic as well as fermionic many-body systems and kinetic equations often occurring in modeling classical flows, have been developed \cite{MendozaPRL2014}. 

Also when solving for a stationary solution it is common to use a time-dependent equation \cite{RuprechtPRA1995} including dissipative damping, or to rewrite the evolution in an unphysical so-called imaginary time; see, e.g., \cite{ChinKrotscheck05}. In this work we will extend this idea by introducing a second-order derivative using an unphysical time parameter. 

In this article we present a new versatile method for solving \eqref{eq:GPE} and other nonlinear equations numerically. The method can be used for systems with extra constraints and more complex nonlinearities. Generalizations of (\ref{eq:GPE}), e.g., with two coupled components \cite{MuellerHo2002}, and even with different types of nonlinearities, as, e.g. quartic for modeling Fermi-Bose mixtures \cite{AbdullaevPRA2013} can also be solved with the method presented here.

\subsection{The dynamical functional particle method}

Let us start quite general and assume that $\mathcal{F}$ is an operator,
$v=v(x),\,v:X\rightarrow\mathbb{R}^{k},\,k\in \mathbb{N}$, and consider the
abstract equation
\begin{equation}
\mathcal{F}(v)=0.\label{Lu}
\end{equation}
In this paper Eq. \eqref{Lu} will be the nonlinear Schr\"{o}dinger
equation. Further, introduce a parameter $\tau$ that belongs to some
(unbounded) interval $T=[0,t_{1}],\,t_{1}\leq\infty$ and define a
new equation in $u=u(x,\tau),\,u:X\times T\rightarrow\mathbb{R}^{k}$
as 
\begin{equation}
\mathcal{M} u_{\tau\tau}+\eta u_{\tau}=\mathcal{F}(u),\label{LuTime}
\end{equation}
where $\mathcal{M}=\mathcal{M}(x,u(x,t),t),\eta=\eta(x,u(x,t),t)$ are parameters.
From physics we recognize (\ref{LuTime}) as a second-order damped
system where $\mathcal{M}$ represents mass and $\eta$ the damping. Together
with the two initial conditions on $u$ we will use (\ref{LuTime})
in such a way that $u_{t},\,u_{tt}\rightarrow0$ when $t\rightarrow t_{1}$,
i.e., $\lim_{t\rightarrow t_{1}}u(x,t)=v(x).$ In other words, we
will solve the damped system (\ref{LuTime}) in order to attain the
stationary solution $v(x).$ For simplicity we use $\mathcal{M}=1$ and $\eta$
constant (chosen to get fast convergence of the dynamical system).

We call the approach for solving (\ref{Lu}) using (\ref{LuTime})
the dynamical functional particle method (DFPM) \cite{Gulliksson}.

\subsection{The model under study}

As a first example we consider the NLSE \eqref{eq:GPE} with an attractive interaction
parameter, corresponding to $U_{0}<0$, on a ring with radius R. Such
systems can model strongly confined BECs in axially symmetric traps
studied experimentally, see e.g. \cite{Guptaetal2005}. This system
has a known analytic solution, described in detail in Appendix \ref{sub:Appendix:A},
which will be used as a reference to the numerical solution. 

We use the domain $-\pi R \leq x < \pi R$ with periodic boundary conditions and introduce the dimensionless angle coordinate $\Theta=x/R$. We divide all terms
in (\ref{eq:GPE}) with $\hbar^{2}/\left(2MR^{2}\right)$
and insert a wave function $\Psi=\varphi/\sqrt{N_\varphi}$ normalized to
unity. We can then introduce a dimensionless coupling constant $\gamma=N_\varphi MRU_{0}/\left(\pi\hbar^{2}\right)$,
such that we obtain the normalized equation

\begin{equation}
-\frac{\partial^{2}\Psi}{\partial\Theta^{2}}+2\pi\gamma\left|\Psi\right|^{2}\Psi=\mu\Psi,\label{eq:GPE_ring}
\end{equation}
where $\mu$ is now a dimensionless eigenvalue. The energy functional corresponding to \eqref{eq:GPE_ring} is 
\begin{equation}
E[\Psi]=\int_{-\pi}^{\pi}\left(\left|\frac{\partial\Psi}{\partial\Theta}\right|^{2}+\gamma\pi\left|\Psi\right|^{4}\right)\mathrm{d}\Theta,\label{eq:Energy_functional}
\end{equation}
subject to the normalization constraint 
\begin{equation}
g_1[\Psi]:=\int_{-\pi}^{\pi}\left|\Psi\right|^{2}\mathrm{d}\Theta-1=0.\label{eq:Normalization}
\end{equation}
Equation (\ref{eq:GPE_ring}) is obtained from the first variation of the constrained energy functional
\begin{equation}
E_{\mu}[\Psi]=E[\Psi]-\mu g_1[\Psi],\label{eq:E_funct_w_norm_constr}
\end{equation}
with respect to $\overline{\Psi}$. 
Variation with respect to $\mu$ gives Eq. \eqref{eq:Normalization}. 

Applying the DFPM \eqref{LuTime} to (\ref{eq:E_funct_w_norm_constr}), keeping the normalization constraint, gives
the damped oscillating system 
\begin{equation}
\frac{\partial^{2}\Psi}{\partial\tau^{2}}+\eta\frac{\partial\Psi}{\partial\tau}=\frac{\partial^{2}\Psi}{\partial\Theta^{2}}-2\pi\gamma\left|\Psi\right|^{2}\Psi+\mu\Psi,\label{eq:DFPM1}
\end{equation}
where, as before, $\eta$ is the damping constant and $\tau$ is a dimensionless time parameter. Note that this
equation is not the time-dependent NLSE. It is an unphysical equation that is
constructed to have a solution that converges to a solution of the time-independent NLSE
(\ref{eq:GPE_ring}). The original functional (\ref{eq:E_funct_w_norm_constr})
is a potential for a damped oscillating system whose stationary
state is the solution of the NLSE (\ref{eq:GPE_ring}).

\section{Numerics, Convergence and Accuracy\label{sec:Convergence,-Accuracy-and}} 

In order to solve (\ref{eq:DFPM1}) numerically we discretize in space and use a numerical method for the resulting system of ordinary differential equations. However, we first rewrite the Eq. (\ref{eq:DFPM1}) as a first-order system where we define the variables $q:=\Psi$ and $p:=\partial \Psi/\partial \tau$.
We then have the dynamical system,
\begin{equation}
\begin{aligned}\dot{q}= & p,\\
\dot{p}= & \frac{\partial^{2}q}{\partial\Theta^{2}} - 2\pi\gamma\left|q\right|^{2}q + \mu q - \eta p.
\end{aligned}
\label{eq:GPE1ord}
\end{equation} 

The unknown eigenvalue $\mu$ can be replaced by the integral $\mu=\int\overline{q}\left(-\frac{\partial^{2}q}{\partial\Theta^{2}}+2\pi\gamma\left|q\right|^{2}q\right)\mathrm{d}\Theta$, as can be seen by multiplying Eq. \eqref{eq:GPE_ring} with $\overline{\Psi}$, integrating over the domain and using the constraint \eqref{eq:Normalization}.

Let $\left\{x^i\right\}_{i=1}^N = \left\{ -\pi,\: -\pi + h, \ldots \pi-h\right\} $, be a partition of the interval $[-\pi,\pi)$ in $N$ points, where $h=2\pi/N$, and let $q_{n}^{i},\:p_{n}^{i}$ represent the values
of $q,\:p$ at the point $x^{i}$ at some time $\tau_{n} = n\vartriangle\!\tau$.
We modify the leapfrog method \cite{HairerLubichWanner} to the damped oscillating system \eqref{eq:DFPM1} 
\begin{equation}
\begin{aligned}p_{n+1/2}^{i}= & p_{n-1/2}^{i}+\vartriangle\!\!\tau\:(F[q_{n}^{i}] + \mu_n q^i_n - \eta\:p_{n-1/2}^{i}),\\
q_{n+1}^{i}= & q_{n}^{i}+\vartriangle\!\!\tau\ p_{n+1/2}^{i},
\end{aligned}
\label{eq:DFPMLeapfrog}
\end{equation}
where 
\begin{equation}
F[q_{n}^{i}] = \dfrac{N}{4\pi}\left(q_{n}^{i+1}-2q_{n}^{i}+q_{n}^{i-1}\right)-2\pi\gamma\left|q_{n}^{i}\right|^{2}q_{n}^{i}, \label{eq:Force}
\end{equation}
and $\mu_n$ is calculated with the trapezoidal approximation of the integral.
The unit norm \eqref{eq:Normalization} is maintained by normalization of ${q_{n}^{i}}$ at each time step. 

We can measure the convergence
of the numerical method to the known analytic solution of the continuous
problem (\ref{eq:MagnusSolution}), sampled at the $N$ grid points,
in the Euclidean norm $\varepsilon_{n} = N^{-1}\sqrt{\sum\limits_{i} \left({q_{n}^{i}-\Psi^{exact}(x^{i})}\right)^2}$.
Running the solver until the error function converges
to a stable minimum $\varepsilon_{n}\rightarrow\varepsilon_{\min}$
for several different values of $N$ gives us an estimate of the dependence
of this minimum on $N$, which is of order $h^2$, as can be seen from Table \ref{tab:convergence}. We use the initial data $q^j_0 = \left(1 + \exp(ix^j)\right)/(\sqrt{2\pi})$, $p_0^j = 0$ for all runs in this section.

\begin{table}
\centering
\begin{ruledtabular}
\begin{tabular}{c|ccccc}
$N$ & 1000 & 2000 & 4000 & 8000 & 16000\tabularnewline
\hline 
$\Delta \tau$ & $6.2 \cdot 10^{-3}$ & $3.1 \cdot 10^{-3}$ & $1.5 \cdot 10^{-3}$ & $7.5 \cdot 10^{-4}$ & $3.8 \cdot 10^{-4}$ \tabularnewline
$\eta$ & 2.6 & 2.6 & 2.6 & 2.6 & 2.6 \tabularnewline
$\varepsilon_{\min}$ & $2.0\cdot10^{-7}$ & $4.1\cdot10^{-8}$ & $9.7\cdot10^{-9}$ & $2.6\cdot10^{-9}$ & $7.8\cdot10^{-10}$\tabularnewline
\end{tabular}
\end{ruledtabular}
\protect\caption[Comparison of discrete and continuous solutions of the NLSE]{\label{tab:convergence}Comparison of discrete and continuous solutions of the NLSE (\ref{eq:GPE_ring}). The table shows the dependence of $\varepsilon_{min}$
on $N$ for the DFPM with $\gamma=-1$. The damping $\eta$ and timestep $\Delta \tau$ are also indicated for each run.}
\end{table}

\subsection{A quantum phase transition} 

As described in Appendix \ref{sub:Appendix:A}, Eq. (\ref{eq:mu_discontinuity}),
there is a discontinuity in the value of the derivative of the chemical
potential when considered as a function of $\gamma$. At $\gamma=-1/2$
the ground-state solution of (\ref{eq:GPE_ring}) undergoes a phase
transformation, from having a localized density profile at lower values
to a completely uniform distribution for larger values of $\gamma$. 

The NLSE is difficult to solve numerically for values of $\gamma$
close to this critical value and it is therefore interesting as a
challenging test for any numerical method. To test how well the numerical method
resolves this discontinuity we solve the discrete equations on a grid
of $N=500$ points, for 81 equidistant values of $\gamma$ between
-0.5100 and -0.4900, for a fixed number of 200 000 iterations for
each $\gamma$, and calculate the central difference approximation for
the derivative of $\mu$ with respect to $\gamma$. 

In order to assess the accuracy of the DFPM we also implement the
so-called ``imaginary time'' method of finding stationary states
to the NLSE. This technique in effect solves the time-dependent NLSE, but for a time variable that takes values on the imaginary
axis, which transforms the dynamics from a wave motion to an exponential
decay of energy to the ground state. In particular we consider the
numerical method using exponential integrators and split operator
techniques, described in \cite{ChinKrotscheck05}, there denoted as
``4A00''. This is, to the best of our knowledge, one of the most efficient numerical
methods previously applied to the NLSE. We have also recently noted a promising method for the stationary and time-dependent NLSE using smoothed-particle hydrodynamics numerical methods \cite{PhysRevE.91.053304} but it is not clear if this method can handle constraints on the equations and it seems less suitable for vortex states as here, where the density can become zero.

\begin{figure}[!h]
\centering
\includegraphics[scale=0.70]{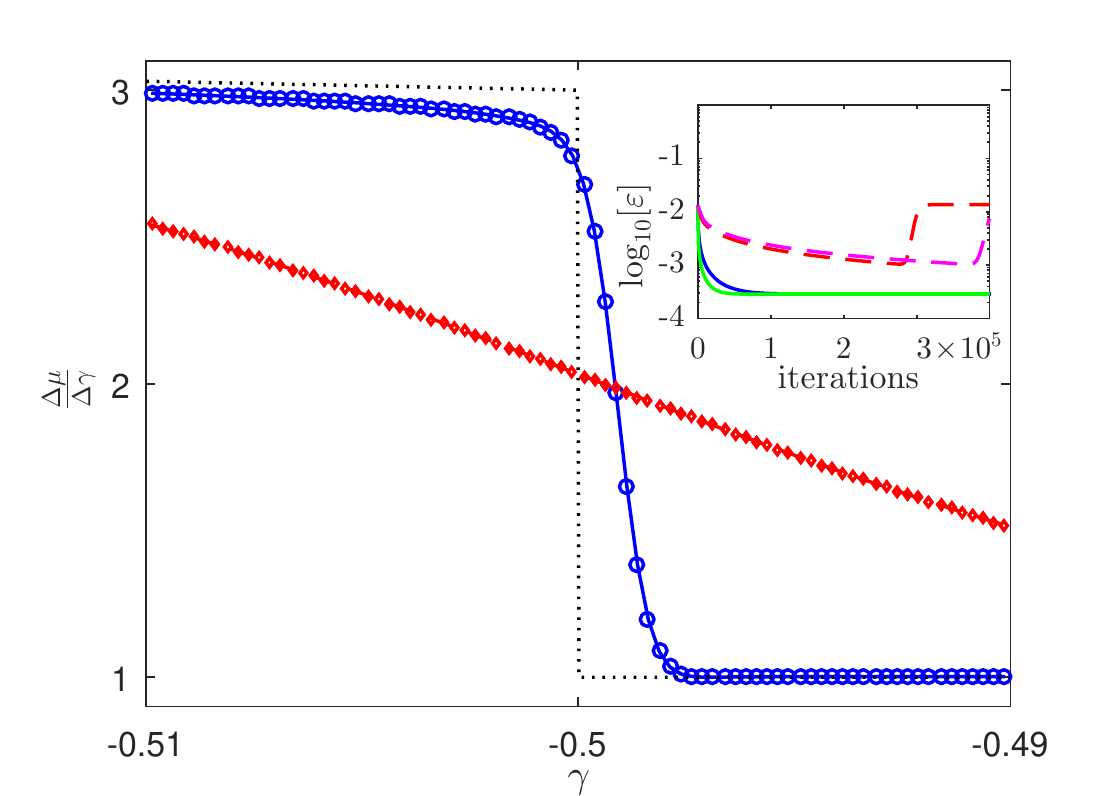}
\protect \caption[Accuracy of the resolution of the discontinuity]{(Color online) Accuracy of the resolution of the discontinuity in
the derivative of $\mu$ with the two different numerical methods. The main figure shows the discontinuity
at $\gamma=-0.5$. DFPM is shown as a solid line with circles
(blue) and 4A00 with diamonds (red). The (black) dotted line is based on the exact solutions
of Appendix \ref{sub:Appendix:A}. The inset plot shows the instability
in $\varepsilon$ of the 4A00-method (dashed) with $\Delta \tau = 0.0002$ (red) and $\Delta \tau = 0.00015$ (magenta) for $\gamma=-0.501$, while the
DFPM (solid) reach a constant $\varepsilon$ both for $\eta = 2.2$ (blue) and $\eta = 1$ (green).}
\label{fig:discontinuity}
\end{figure}

In the tests we used the parameters $\eta=2.2$, $\Delta \tau = 0.012$ for the DFPM, and the imaginary time implementation 4A00 used a time
step of $\Delta \tau = 0.0002$. These values of $\Delta \tau$ were the largest we could
find that were stable for this value of $N$ during 200 000 iterations for the respective methods. The resulting value for the derivative is plotted in Fig. \ref{fig:discontinuity}
together with the exact values calculated numerically from the expression for $\mu$
given in (\ref{eq:mu_ref}) of Appendix \ref{sub:Appendix:A}.

Both methods converge slower close to the discontinuity and give less
accurate solutions as well as chemical potentials. However, the
DFPM produces a qualitatively correct step (see Fig. \ref{fig:discontinuity}),
while the 4A00 method performs less well here. In fact it is even
worse than Fig. \ref{fig:discontinuity} shows. If we let the numerical
solvers continue until they stabilize we find that 4A00 diverges; see
the inset of Fig. \ref{fig:discontinuity} for two examples. We have
in fact been unable to find any value of $\Delta \tau$ for which 4A00 stabilizes to
a value close to the exact solutions for the range of $\gamma$ values investigated. The DFPM may also fail if the time step is too large but it is always possible to find some $\Delta \tau$ such that convergence is assured. 

Changing the damping in the DFPM does affect the speed of convergence somewhat but does not affect the end result as long as the solver is allowed to stabilize. The inset of Fig. \ref{fig:discontinuity} also shows the convergence of DFPM for two different values of the damping parameter: $\eta = 1.0,\ 2.2$. 

\section{Constraints on the solutions\label{sec:Constrains-for-the}} 

Ground-state solutions to the NLSE with a nonzero angular momentum
are called yrast states. These solutions can be considered as stationary
when viewed from a co-rotating frame. If we introduce a new angle
coordinate $\theta=\Theta-\Omega\tau$ to the time-dependent NLSE
corresponding to (\ref{eq:GPE_ring})

\begin{equation}
i\frac{\partial\Psi}{\partial\tau} = -\frac{\partial^{2}\Psi}{\partial\Theta^{2}} + 2\pi\gamma\left|\Psi\right|^{2}\Psi, \label{eq:TDNLSE}
\end{equation}
and use the ansatz $\Psi\left(\theta\right)=\Psi\left(\Theta\right)\exp\left(i\Omega\tau\right)$,
we obtain from (\ref{eq:TDNLSE}) the following equation in the coordinate
$\theta$:
\begin{equation}
- \frac{\partial^{2}\Psi}{\partial\theta^{2}} + i\Omega\frac{\partial\Psi}{\partial\theta} + 2\pi\gamma\left|\Psi\right|^{2}\Psi = \mu\Psi.\label{eq:NLSE_in_R_F}
\end{equation}
The corresponding angular momentum is given by the functional,
\begin{equation}
\ell=-i\int_{-\pi}^{\pi}\overline{\Psi}\frac{\partial\Psi}{\partial\theta}\mathrm{d}\theta.\label{eq:Angular_momentum}
\end{equation}

Alternatively to the time-dependent NLSE (\ref{eq:TDNLSE}) we can
introduce two Lagrange multipliers corresponding to the chemical
potential and the angular velocity, respectively, $\lambda^1 := \mu$ for the normalization
constraint \eqref{eq:Normalization}, and
$\lambda^2 := \Omega$ for the additional constraint,
\begin{equation}
g_2[\Psi] := -i\int_{-\pi}^{\pi}\overline{\Psi}\frac{\partial\Psi}{\partial\theta}\mathrm{d}\theta - \ell_0 = 0,
\end{equation}
for some fixed value $\ell_0$ of the angular momentum. Minimizing the following functional, 

\begin{equation}
E_{\lambda}\left[\Psi\right] = E[\Psi] - \lambda^Ag_A[\Psi],\label{eq:Functional_with_constraints}
\end{equation}
is then equivalent to finding the ground-state solution to (\ref{eq:TDNLSE}),
since (\ref{eq:NLSE_in_R_F}) is the corresponding Euler equation
to (\ref{eq:Functional_with_constraints}). Einstein's summation convention applies to all index pairs that appears both as superscript and subscript.

The DFPM can be used for problems with constraints straightforwardly. As before, instead
of solving (\ref{eq:NLSE_in_R_F}) directly, we consider the time-dependent unphysical
problem, 
\begin{equation}
\frac{\partial^{2}\Psi}{\partial\tau^{2}}+\eta\frac{\partial\Psi}{\partial\tau}= F[\Psi] + \lambda^AG_A[\Psi],\label{eq:DFPM_for_NLSE_constr}
\end{equation}
where $F[\Psi] := - \delta E[\Psi]/\delta\overline{\Psi}$, $G_A[\Psi]:=\delta g_A[\Psi]/\delta \overline{\Psi}$ are defined by functional derivation of the energy and constraint functionals, respectively.

We have chosen to solve \eqref{eq:DFPM_for_NLSE_constr} by a modified version of the second-order symplectic method RATTLE \cite{RATTLE}, originally considered to handle separable Hamiltonians with constraints. As a first step we once again rewrite \eqref{eq:DFPM_for_NLSE_constr} as a first-order system similar to \eqref{eq:GPE1ord} 
\begin{equation}
\begin{aligned}\dot{q}= & p,\\
\dot{p}= & F[q] + \lambda^{A}G_A[q] - \eta p, \quad A=1,2.
\end{aligned}
\label{eq:DFPM_constr_firstord}
\end{equation}

To solve this system numerically we discretize in space as before and use the RATTLE method for the time evolution, with the modification that the nonsymplectic part of \eqref{eq:DFPM_constr_firstord} modifies the update of the momentum according to  

\begin{equation}
\begin{aligned}p^i_{n+1/2}= & p^i_{n}+\dfrac{\Delta \tau}{2}\left(F[q^i_n] + \lambda_{n}^{A}G_A[q^i_n] - \eta p^i_{n}\right),\\
q^i_{n+1}= & q^i_{n}+\Delta \tau p^i_{n + 1/2},\\
0 = & g_A[q^i_{n+1}],\quad A=1,2,\\
p^i_{n+1}= & p^i_{n+1/2}+\dfrac{\Delta \tau}{2}\left(F[q^i_{n+1}] + \tilde{\lambda}_{n}^{A}G_A[q^i_{n+1}] - \eta p^i_{n+1/2}\right),\\
0= & \int_{-\pi}^{\pi} \overline{G_A[q_{n+1}]} p^i_{n+1} \mathrm{d}\theta,\quad A=1,2.
\end{aligned}
\label{eq:DFPMRATTLE}
\end{equation}

The last equation above is a projection step to ensure that the update of
$p^i_n$ is tangent to the constraint surface. The integral is performed with the trapezoidal approximation. The constraints $g_A[q^i_{n+1}]$
are quadratic algebraic equations in the Lagrange multipliers $\lambda_{n}^{A}$,
and are solved with Newton's method for each time step $n$. The linear
projection equation is solved for a second set of Lagrange multipliers
$\tilde{\lambda}_{n}^{A}$ (see Chapter VII of \cite{HairerLubichWanner})
that coincide with $\lambda_{n}^{A}$ in the limit of the stationary
solution. 

\begin{figure}[!ht]
\centering
\includegraphics[scale=0.424]{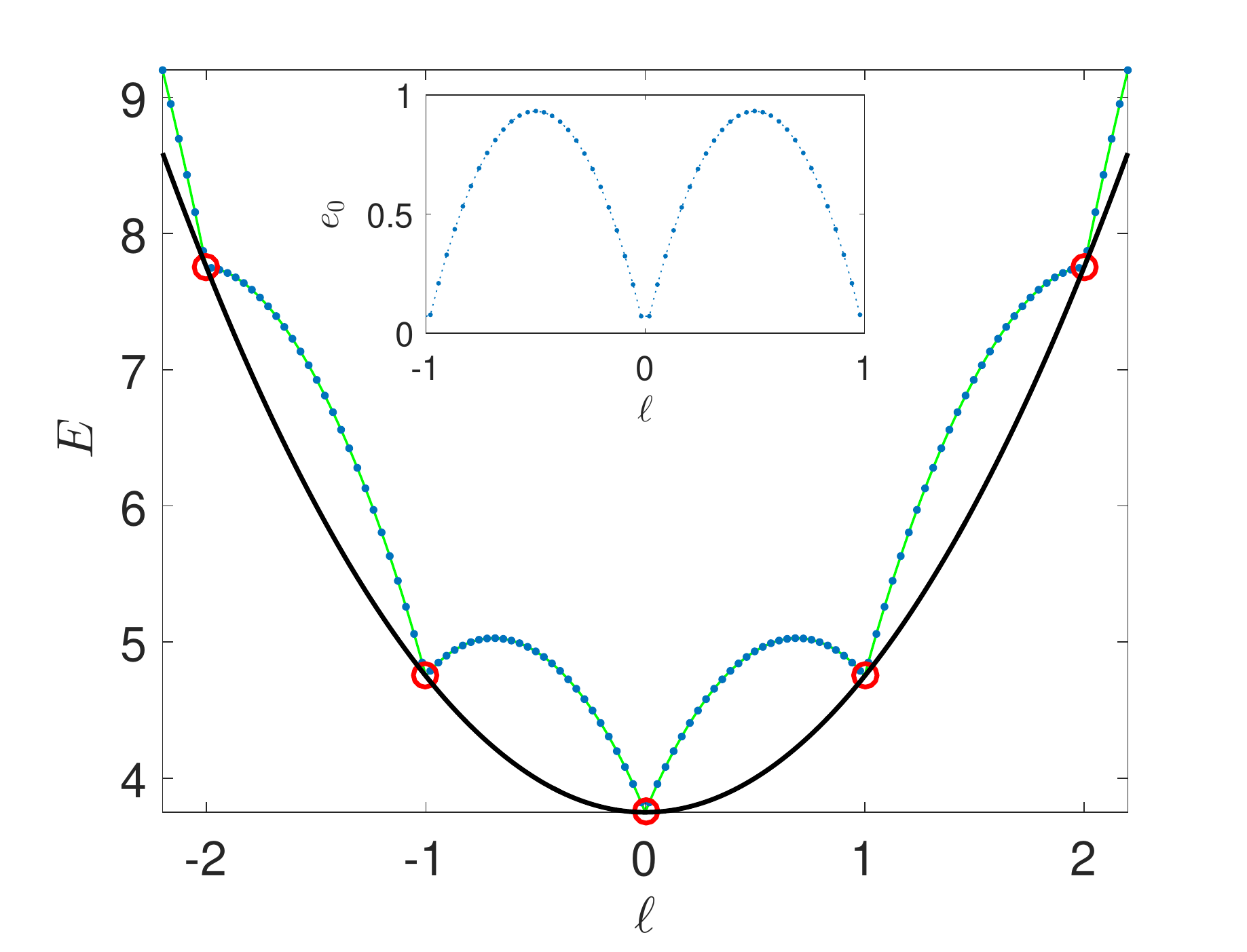}
\protect\caption[Yrast curve for the single-component NLSE]{(Color online) Yrast curve for the single-component NLSE with $\gamma = 7.5$. The figure compares the numerical results with the values derived from the analytic solutions given in Appendix \ref{sub:Appendix:B} and verifies Bloch's formula \eqref{eq:Bloch} numerically. The (blue) dots show numerical results from solving Eqs. (\ref{eq:NLSE_in_R_F}) and (\ref{eq:Angular_momentum}) with the DFPM \eqref{eq:DFPMRATTLE}. The thin (green) solid line shows the yrast curve obtained from the analytic results given in Appendix \ref{sub:Appendix:B}. The energy of the plane-wave states \eqref{eq:pwstates} are indicated as (red) circles. The quadratic $\ell$ dependence for the energy is shown as a thick solid (black) line and the numerical realization of the periodic function $e_{0}(\ell)$ is obtained by subtracting the black curve from the main numerical data and is plotted in the inset figure with (blue) dots connected by a dotted line.}
\label{fig1} 
\end{figure}

Our goal here is to obtain $E$ from (\ref{eq:Energy_functional})
as a function of the angular momentum $\ell$ that is given by the
functional (\ref{eq:Angular_momentum}). Hence, given a fixed value for
the angular momentum, $\ell_0$, we solve the constraints and Eq. (\ref{eq:NLSE_in_R_F})
for $\mu,\:\Omega$ and $\Psi$. Having found this $\Psi$, the energy
$E$ is calculated from (\ref{eq:Energy_functional}) and we plot
the so-called yrast curve $E \left( \ell \right)$ (see Fig. \ref{fig1}). 

We used a grid of $N=400$ points, a damping parameter $\eta = 2.74$, and time step $\Delta \tau = 0.015$. Initial data were chosen for each run as $q^j_0 = \left(a + \sqrt{\ell_0/k}\exp(ikx^j) \right) /(\sqrt{2\pi})$, $p^j_0 = 0$, where $k$ is the nearest integer with absolute value larger than or equal  to $| \ell_0 |$ and $a$ is chosen such that the normalization is equal to 1. These initial data satisfied both constraints and produced good results for all $| \ell_0 | < 1$ but not for larger values. For these we insted used a $k$ that had an absolute value that was the next larger integer.

In Appendix \ref{sub:Appendix:B} we outline how to implicitly represent the yrast curve in terms of elliptic integrals and Jacobi elliptic functions, that we have used to benchmark the numerical results.
At integer values
of the angular momentum, the ground-state solutions to the NLSE are
the plane-wave states, 
\begin{equation}\label{eq:pwstates}
\varphi_{k}(\theta)=\exp(ik\theta)/\sqrt{2\pi}.
\end{equation}
As a check of the numerical results we also plot the energy of these
states in Fig. \ref{fig1}.

According to Bloch's theorem \cite{BlochPRA1973}, the energy can
be split into a constant part, a part with quadratic dependence on
$\ell$, and a periodic and symmetric part, 
\begin{equation}\label{eq:Bloch}
E(\ell)=\gamma/2 + \ell^{2} + e_{0}(\ell).
\end{equation}
Both the periodicity and symmetry of $e_{0}(\ell)$ is verified numerically as seen
in the inset of Fig. \ref{fig1}.

The solution to the DFPM preserves the constraints numerically to a given set tolerance
at each time step and converges to an approximation of order $h^2$ to the solution of the continuous NLSE. An alternative method often used for solving the constrained NLSE is the so-called penalty method~\cite{PhysRevA.72.053624} where the functional, 
\begin{equation}
E_{w}\left[\Psi\right]=E\left[\Psi\right]+\frac{w}{2}\left(g_2[\Psi]\right)^{2},\label{eq:Penalty_method}
\end{equation}
is minimized for a constant weight $w$. The drawback with the penalty method
is that the minimum of this functional is not the minimum of $E\left[\Psi\right]$
and that the momentum of the solution will not be $\ell_0$, but
some value close to $\ell_0$. The weight $w$ in (\ref{eq:Penalty_method})
has to be chosen such as to balance the error in the original energy functional
with the error in the angular momentum constraint, since they can
not attain minimal value at the same time. The DFPM presented has none of these
drawbacks.

\section{Constraints for Two-component systems} 

Experiments on persistent currents in toroidal two-component Bose
gases \cite{Beattieetal2013} have motivated theoretical investigation
of the coupled multi component NLSE on a ring geometry \cite{WuZaremba2013, Smyrnakisetal2014}.
The multi component NLSE can be solved numerically with the DFPM with
only minor modification from the constrained single-component case
if we consider the two species as components of a vector-valued mean-field
wave function. The main difference here is that we now have three
constraints instead of two, and that the nonlinearity parameter becomes
a matrix-valued coupling between the different components. 

Let $\Psi_{s},\:s=1,2$ denote the mean-field wave functions of the
two components. Each component satisfies a normalization constraint
$\int\overline{\Psi_{s}}\Psi_{s}\mathrm{d}\theta=x_{s}$, where the total
normalization sums up to unity, $x_{1}+x_{2}=1$. The yrast states
are obtained using the further constraint that the total angular momentum
is constant, $-i\int\left(\overline{\Psi}_{1}\frac{\partial\Psi_{1}}{\partial\theta}+\overline{\Psi}_{2}\frac{\partial\Psi_{2}}{\partial\theta}\right)\mathrm{d}\theta=\ell_0$.
All three constraints can be written on the general form, 
\begin{equation}
\int_{-\pi}^{\pi}\pmb{\Psi}^{\dagger}\pmb{K}_{A}\pmb{\Psi} \mathrm{d}\theta=c_{A},
\label{eq:2compconstraints}
\end{equation}
for $(c_A) = (x_1,\: x_2, \: \ell_0)$, and three different matrices of operators,
\begin{equation}
\pmb{K}_{1}=\begin{bmatrix} {\rm I} & 0\\
0 & 0
\end{bmatrix},\:\pmb{K}_{2}=\begin{bmatrix}0 & 0\\
0 & {\rm I}
\end{bmatrix},\:\pmb{K}_{3}=-i\begin{bmatrix}\frac{\partial}{\partial\theta} & 0\\
0 & \frac{\partial}{\partial\theta}
\end{bmatrix},
\end{equation}
where I denotes the identity map, acting on $\pmb{\Psi}$, with the conjugate transpose $\pmb{\Psi}^{\dagger}=\begin{bmatrix}\overline{\Psi}_{1} & \overline{\Psi}_{2}\end{bmatrix}$.
The problem is then to find the minimum of the two-component energy-functional
\begin{equation}
E[\pmb{\Psi}] = \int_{-\pi}^{\pi}\pmb{\Psi}^{\dagger}\left(-\frac{\partial^{2}}{\partial\theta^{2}}+\pi\pmb{\Gamma}_{\Psi}\right)\pmb{\Psi} \mathrm{d}\theta,\label{eq:two_comp_energy}
\end{equation}
subject to the constraints, which are added to the energy-functional
together with the triplet of Lagrange multipliers $\lambda^{A}=(\mu_{\text{1}},\mu_{2},\Omega)$,
resulting in the constrained energy functional, 
\begin{eqnarray}
E_{\lambda}[\pmb{\Psi}] & = E[\pmb{\Psi}] + \lambda^{A}\left(c_{A}-\int_{-\pi}^{\pi}\pmb{\Psi}^{\dagger}\pmb{K}_{A}\pmb{\Psi} \mathrm{d}\theta\right),\label{eq:TwoCompEnergy}
\end{eqnarray}
where 
\begin{equation}
\pmb{\Gamma}_{\pmb{\Psi}}=\begin{bmatrix}\gamma_{11}|\Psi_{1}|^{2} & \gamma_{12}\Psi_{1}\overline{\Psi}_{2}\\
\gamma_{21} \Psi_2 \overline{\Psi}_{1} & \gamma_{22}|\Psi_{2}|^{2}
\end{bmatrix}.
\end{equation}
Variation with respect to $\pmb{\Psi}^{\dagger}$ gives the corresponding coupled
NLSEs,

\begin{equation}
-\frac{\partial^{2}\pmb{\Psi}}{\partial\theta^{2}}+2\pi\pmb{\Gamma}_{\pmb{\Psi}}\pmb{\Psi}-\lambda^{A}\pmb{K}_{A}\pmb{\Psi}=0.\label{eq:2compNLSE}
\end{equation}

Using a discretization of $N=400$ points, i.e., a state vector $q^i_n$ with $2N$ points, representing both components, we can use the DFPM formulation (\ref{eq:DFPMRATTLE}) to solve
Eq. (\ref{eq:2compNLSE}). 

\begin{figure}[!ht]
\centering
\includegraphics[scale=0.42]{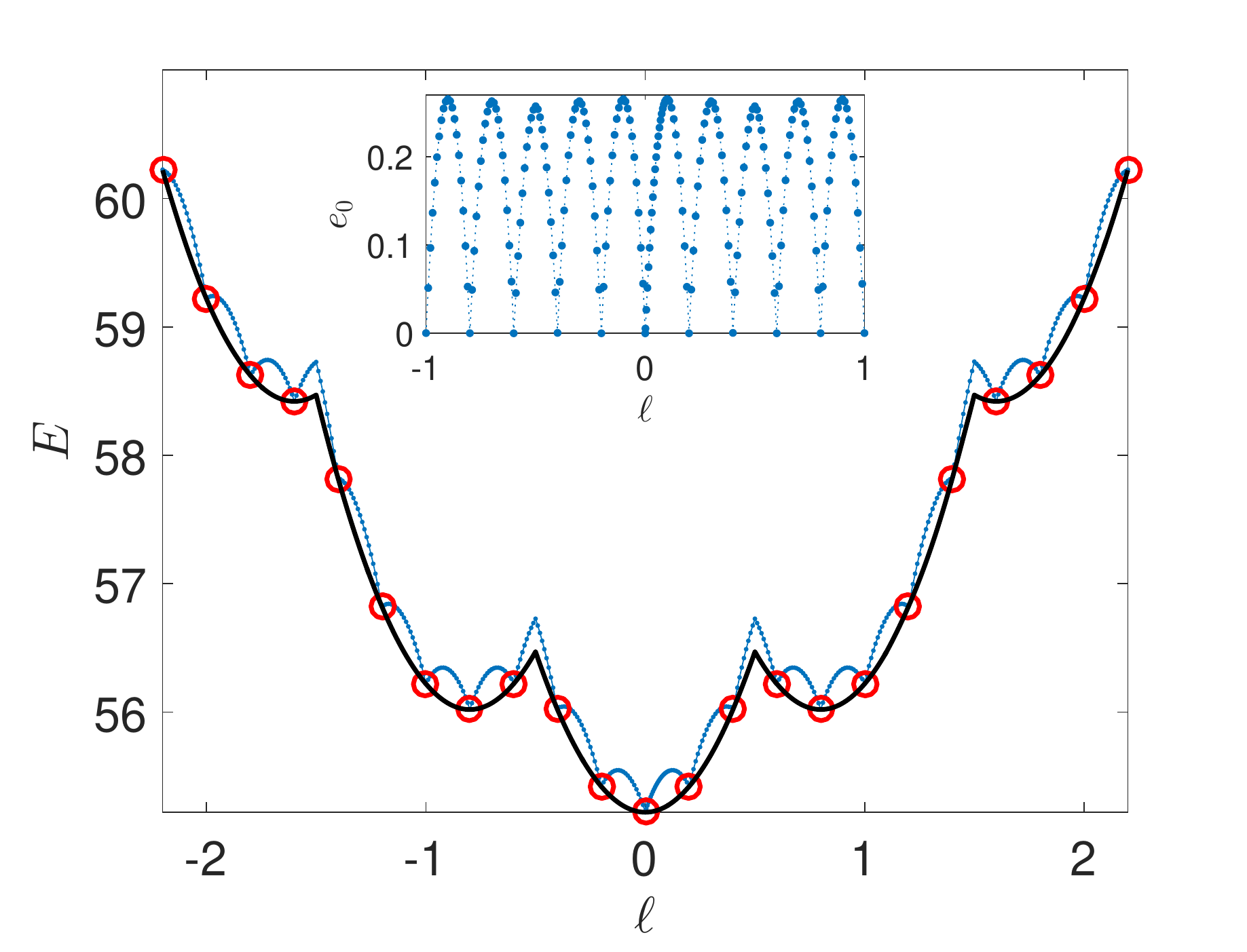}
\protect\caption[Yrast curve for the two-component
NLSE]{(Color online) Yrast curve for the two-component
NLSE. The figure is the two-component analog to Fig. \ref{fig1}. 
The main figure shows both numerical results obtained using the algorithm 
\eqref{eq:DFPMRATTLE}, indicated by (blue) dots connected by a thin solid line to 
guide the eye, and analytic results from the plane-wave solutions of Eq. 
(\ref{eq:pwEnergies}), indicated as (red) circles. In addition, the nonperiodic 
component of the function \eqref{eq:Bloch2comp}, $ E_{\mathrm{int}} + P_{0}(\ell) $ 
is plotted as a thick (black) solid line. The inset figure shows the numerical data 
for the periodic part of the yrast curve, $e_{0}$, which is obtained by subtracting 
the black curve from the numerical data in the main figure.}
\label{fig:2compyrast}
\end{figure}

As an example, we use the parameter values $x_{1}=0.8,\:x_{2}=0.2,\:\gamma_{11}=\gamma_{22}=1250/\pi^{2},\:\gamma_{12}=\gamma_{21}=750/\pi^{2}$,
and solve Eq. (\ref{eq:2compNLSE}) for 460 values of $\ell$
between -2.2 and 2.2. The resulting yrast curve is shown in Fig. \ref{fig:2compyrast}. The equations were solved for $\ell$ in sequence, where initial data for the next run was given by the solution from the last. The damping parameter was initially set to 1. Sometimes the solver gave solutions with too high a value for the energy and then the damping parameter was halved and the solver restarted with the lower value for the damping until a solution with an energy close to the last point on the curve (presumably being the yrast state) was found.

As a check of the results we also plot the energies of the analytic two-component
plane-wave states that should lie on the yrast curve as local minima
of the energy for the parameters chosen here \cite{Smyrnakisetal2014}. The plane-wave states are given
by $\Psi_{s}(\theta)=\sqrt{x_{s}/2\pi}\exp(ik_{s}\theta)$,
and the total energy and angular momentum of these states are according to \eqref{eq:2compconstraints} and \eqref{eq:two_comp_energy} 
\begin{align}
E & =x_{1}k_{1}^{2}+x_{2}k_{2}^{2}+ E_{\mathrm{int}},\label{eq:pwEnergies}\\
\ell & =x_{1}k_{1}+x_{2}k_{2},\nonumber 
\end{align}
with $E_{\mathrm{int}} = \left( \gamma_{11}x_{1}^{2} +\gamma_{22} x_{2}^{2}  + \gamma_{12} x_{1}x_{2} +\gamma_{21} x_{1}x_{2} \right)/2$. We have plotted $E(\ell)$ of (\ref{eq:pwEnergies}) as circles in
the yrast curve for a subset of integer wave numbers $k_{s}$ between
-3 and 3. The numerical results of the DFPM coincide with the analytic
solutions of (\ref{eq:pwEnergies}) at the points where the plane-wave
solutions are applicable (see Fig. \ref{fig:2compyrast}).

As illustrated in Fig. \ref{fig1}, the energy of the single-component NLSE yrast state can be split into
one part with quadratic dependence of the angular momentum and one
part that is periodic \cite{BlochPRA1973}. One can see from Fig.
\ref{fig:2compyrast} that the two-component yrast curve has two
different quadratic energy scales, corresponding to the majority-
($x_{1}=0.8$) component and the minority- ($x_{2}=0.2$) component,
respectively. There is a major quadratic dependence $P_{1}(\ell)=\ell^{2}/x_{1}$,
and superimposed on this function there are smaller parabolas $P_{2}^{n}(\ell)=(\ell-nx_{1})^{2}/x_{2},\:n\in\mathbb{Z}$.
The yrast curve is determined by the lowest energy value of these
quadratic functions plus a part, $e_{0}(\ell)$, that is periodic and symmetric,
\begin{equation}\label{eq:Bloch2comp}
E(\ell) =  E_{\mathrm{int}} + P_0(\ell) + e_0(\ell),
\end{equation}
where
\begin{eqnarray}
P_{0}(\ell) & = & \underset{n\in\mathbb{Z}}{\min}\left\{ P_{1}(nx_{1})+P_{2}^{n}(\ell)\right\} \label{eq:energyparabolas} \\
 & = &  \left[ \ell \right]^{2}x_{1}+(\ell-\left[ \ell \right]x_{1})^{2}/x_{2} \nonumber.
\end{eqnarray}
Here $\left[\ \right]$ denotes the nearest integer function. Subtracting
the function $E_{\mathrm{int}} + P_{0}(\ell) $ from the numerical data of the energy
gives the periodic function $e_{0}(\ell)$ (see inset of Fig. \ref{fig:2compyrast}).

\section{Conclusions}

We have validated the dynamical functional particle method (DFPM) numerically for retrieving stationary solutions of the nonlinear Schrödinger equation \eqref{eq:GPE}.
With an attractive (negative) interaction parameter, the method managed well in resolving a quantum phase transition, which seems difficult with other numerical methods, and could reproduce analytic results in the limit of an increasing numbers of grid points. 
For a repulsive (positive) interaction parameter we added a constraint on the angular momentum, which allows for nontrivial solutions, and reproduced the so-called yrast curve numerically up to machine precision.
Finally, we added a second component together with a constraint on the total angular momentum, for which we calculated a corresponding yrast curve. 

The method we have  developed  can be generalized in dimensionality, in the number of components, and in the number of and type of constraints. Hence, the  method may be used in a wide range of future applications.

\section{Acknowledgments}

The authors acknowledge support from \"{O}rebro University, School of
Science and Technology, and partly through RR 2015/2016. We thank
G. M. Kavoulakis for useful discussions.

\appendix
\section*{Appendices: Reference solutions for the benchmarking of the numerical
method}

We here give the analytic formulas used as reference solutions for
the numerical simulations presented in Secs. \ref{sec:Convergence,-Accuracy-and}
and \ref{sec:Constrains-for-the} of the article.

\section{The NLSE on a ring with attractive interaction\label{sub:Appendix:A}}

The NLSE presented in Eq. (\ref{eq:GPE_ring}) gives rise to
a quantum phase transition for a critical negative value of the parameter
$\gamma$ in the nonlinear term. For $\gamma\geq-1/2$ the density
is uniform, while for $\gamma<-1/2$ a peak develops in the ground-state density \cite{KanamotoPRA2003,KavoulakisPRA2003}. For $\gamma<-1/2$
the wave function can be expressed as \cite{CarrPRA2000}

\begin{equation}
\Psi=\sqrt{\frac{\textrm{K}\left(m\right)}{2\pi\textrm{E}\left(m\right)}}\textrm{dn}\left(\frac{\textrm{K}\left(m\right)}{\pi}\Theta,m\right),\label{eq:MagnusSolution}
\end{equation}
where $\textrm{K}$ and $\textrm{E}$ are the complete elliptic integrals
of the first and second kind, and $\textrm{dn}$ is a Jacobi elliptic
function \cite{Abramowitz_and_Stegun1965}. In order to chose the
dimensionless parameter $0\leq m<1$ for a given value of $\gamma<-1/2$,
the following equation is solved \cite{KanamotoPRA2003}

\begin{equation}
\textrm{K}\left(m\right)\textrm{E}\left(m\right)=-\frac{\pi^{2}\gamma}{2}.\label{eq:m_gamma_relation}
\end{equation}
As we can see from a power series expansion in $m$ of the elliptic integrals,
$\textrm{K}\left(m\right)\sim\pi/2\left(1+m/4+9m^{2}/64+...\right)$
and $\textrm{E}\left(m\right)\sim\pi/2\left(1-m/4-3m^{2}/64+...\right)$,
the critical coupling $\gamma\rightarrow-1/2^{-}$ corresponds to
$m\rightarrow0^{+}$. We can then relate $\gamma$ and $m$ in this
limit from (\ref{eq:m_gamma_relation}) according to

\begin{equation}
\frac{dm}{d\gamma}=\frac{-4}{\sqrt{-\gamma-1/2}}.\label{eq:dm_div_dgamma}
\end{equation}
The derivative of the chemical potential, i.e., the lowest eigenvalue
of Eq. (\ref{eq:GPE_ring}) which according to \cite{KanamotoPRA2003} is

\begin{equation}
\mu=\left\{ \begin{array}{ll}
\gamma, & \gamma\geq-1/2\\
-\frac{\textrm{K}^{2}\left(m\right)\left(2-m\right)}{\pi^{2}}, & \gamma<-1/2
\end{array}\right.,\label{eq:mu_ref}
\end{equation}
with respect to the parameter $\gamma$, is discontinuous at the critical
value $\gamma=-1/2$ and can there be expressed with the help of (\ref{eq:dm_div_dgamma})
according to

\begin{equation}
\frac{\partial\mu}{\partial\gamma}=\frac{dm}{d\gamma}\frac{d\mu}{dm}\rightarrow\left\{ \begin{array}{ll}
1, & \gamma\rightarrow-1/2^{+}\\
3, & \gamma\rightarrow-1/2^{-}
\end{array}\right..\label{eq:mu_discontinuity}
\end{equation}
This explains the step seen in Fig. \ref{fig:discontinuity}, which
is a critical test for the accuracy of a numerical method.

\section{The NLSE on a ring with repulsive interaction and a constrained
angular momentum\label{sub:Appendix:B}}

The density of the ground state is always uniform for a positive parameter
$\gamma$ in the nonlinear term of Eq. (\ref{eq:GPE_ring}).
However, exciting the ring system to a constrained value of the (normalized)
angular momentum $0<\ell<1$ form gray solitary waves with a nonuniform
density in the rotating frame \cite{CarrPRA2000_I,SmyrnakisPRA2010}.
It was recently pointed out that those solitary waves are indeed the
lowest rotational excitations discussed in the concept of the so called
yrast curve, i.e., the states with the lowest energy given an angular
momentum $0<\ell<1$ \cite{KanamotoPRA2010,JacksonEPL2011}. In order
to benchmark the numerical simulations for the constrained NLSE presented
in Sec. \ref{sec:Constrains-for-the}, we have compared with an
alternative representation of the yrast curve which we have based
on the work presented in \cite{SmyrnakisPRA2010,JacksonEPL2011}.
We use again a dimensionless parameter $0\leq m<1$ but now to parametrize
the angular momentum $\ell\left(m\right)$ and the lowest energy $E\left(m\right)$
given the constraint, in order to plot the yrast curve with $E$ versus
$\ell$, see Fig. \ref{fig1}. With an ansatz $\Psi=\sqrt{n}\exp\left(i\phi\right)$
for the complex wave function with density $n\left(m\right)=\left|\Psi\right|^{2}$
(normalized to unity) and a phase $\phi\left(m\right)$, the normalized
angular momentum (\ref{eq:Angular_momentum}) can be written as

\begin{equation}
\ell\left(m\right)=\int_{-\pi}^{\pi}n\frac{\partial\phi}{\partial\theta}\mathrm{d}\theta.\label{eq:ana_ell}
\end{equation}
Above we have used that $n\left(\theta\right)$ is an even function,
such that the integral of its derivative disappears. For the normalized
energy (\ref{eq:Energy_functional}) we have

\begin{equation}
E\left(m\right)=\int_{-\pi}^{\pi}\left[\left(\frac{\partial\sqrt{n}}{\partial\theta}\right)^{2}+n\left(\frac{\partial\phi}{\partial\theta}\right)^{2}+\pi\gamma n^{2}\right]\mathrm{d}\theta,\label{eq:ana_E}
\end{equation}
where we have used that $1/\left(2\sqrt{n}\right)\partial n/\partial\theta=\partial\sqrt{n}/\partial\theta$.
Below we give our parametrizations of the density, 

\begin{equation}
n\left(m\right)=\frac{1}{2\pi}-\frac{\textrm{K}\left[\textrm{K}-\textrm{E}-m\textrm{K}\textrm{sn}^{2}\left(\frac{\textrm{K}}{\pi}\theta,m\right)\right]}{\pi^{3}\gamma},\label{eq:ana_n_m}
\end{equation}
and the phase 

\begin{equation}
\frac{\partial\phi}{\partial\theta}\left(m\right)=\frac{1}{8\pi^{5}\gamma}\sqrt{abc}\left[\int_{-\pi}^{\pi}\frac{1}{n\left(m\right)}\mathrm{d}\theta-\frac{2\pi}{n\left(m\right)}\right],\label{eq:ana_dphi_dtheta_m}
\end{equation}

\[
\begin{array}{l}
a=4m\textrm{K}^{2}+4\textrm{E}\textrm{K}-4\textrm{K}^{2}+2\pi^{2}\gamma,\\
b=2\textrm{E}\textrm{K}-2\textrm{K}^{2}+\pi^{2}\gamma,\\
c=2\textrm{E}\textrm{K}+\pi^{2}\gamma,
\end{array}
\]
which are needed above in Eqs. (\ref{eq:ana_ell}) and (\ref{eq:ana_E})
in order to produce the yrast curve for a given $\gamma-$parameter.
Again $\textrm{K}$ and $\textrm{E}$ are the complete elliptic integrals
of the first and second kind, and $\textrm{sn}$ is a Jacobi elliptic
function with derivative $\partial\textrm{sn}\left(\theta,m\right)/\partial\theta=\textrm{cn}\left(\theta,m\right)\textrm{dn}\left(\theta,m\right)$
\cite{Abramowitz_and_Stegun1965}, such that $\partial\sqrt{n}/\partial\theta\left(m\right)=\textrm{K}^{3}m\:\textrm{sn}\:\textrm{cn}\:\textrm{dn}/\left(\pi^{4}\gamma\sqrt{n\left(m\right)}\right)$.
Note that here the dimensionless parameter $0\leq m<1$ can be chosen
arbitrarily in order to cover the range $0\leq\ell<1/2$, while in
practice one needs a non-equidistant domain of $m$-values to produce
an yrast curve that is equidistant in the $\ell$ variable. The continuation
of the yrast curve to $1/2<\ell\leq1$ can then be obtained directly
with so-called Bloch mapping \cite{BlochPRA1973}. In a similar way
we can get even the full range $-\infty<\ell<\infty$ (see Fig.
\ref{fig1}).

\bibliography{Bibliography}

\end{document}